\documentclass[trackchanges,twocolumn]{aastex701}

\newcommand{\kms}{\mathrm{km\,s}^{-1}}

\usepackage{amsmath}	% Advanced maths commands
\usepackage{amssymb}	% Extra maths symbols

\begin{document}

\title{A Planetary Nebula from a 5.7\,$M_{\odot}$ Progenitor in a 90\,Myr M31 Star Cluster}

\author[orcid=0009-0007-5623-2475]{Pinjian Chen}
\affiliation{National Astronomical Observatories, Chinese Academy of Sciences, Beijing 100101, P.\,R.\,China}
\affiliation{School of Astronomy and Space Science, University of the Chinese Academy of Sciences, Beijing, 100049, P.\,R.\,China}
\email{} 

\author[orcid=0000-0003-2472-4903]{Bingqiu Chen}
\affiliation{South-Western Institute for Astronomy Research, Yunnan University, Kunming, Yunnan 650091, P.\,R.\,China}
\email[show]{bchen@ynu.edu.cn} 

\author[orcid=0000-0001-7084-0484]{Xiaodian Chen}
\affiliation{National Astronomical Observatories, Chinese Academy of Sciences, Beijing 100101, P.\,R.\,China}
\affiliation{School of Astronomy and Space Science, University of the Chinese Academy of Sciences, Beijing, 100049, P.\,R.\,China}
\affiliation{Institute for Frontiers in Astronomy and Astrophysics, Beijing Normal University, Beijing 102206, P.\,R.\,China}
\email[show]{chenxiaodian@nao.cas.cn} 

\author[orcid=0000-0003-4489-9794]{Shu Wang}
\affiliation{National Astronomical Observatories, Chinese Academy of Sciences, Beijing 100101, P.\,R.\,China}
\affiliation{School of Astronomy and Space Science, University of the Chinese Academy of Sciences, Beijing, 100049, P.\,R.\,China}
\email{}

\author[orcid=0000-0003-2471-2363]{Haibo Yuan}
\affiliation{School of Physics and Astronomy, Beijing Normal University, Beijing 100875, P.\,R.\,China}
\affiliation{Institute for Frontiers in Astronomy and Astrophysics, Beijing Normal University, Beijing 102206, P.\,R.\,China}
\email{}

\begin{abstract}
Planetary nebulae (PNe) trace the late evolution of low-to-intermediate-mass stars, yet the masses of their progenitors are rarely measured directly. Here we present a PN physically associated with a young star cluster in M31, providing an unprecedented extragalactic empirical anchor in the poorly constrained high-mass regime of PN progenitors. High-resolution \textit{Hubble Space Telescope} imaging shows that the nebula lies near the cluster center, and spectral decomposition of the blended cluster-plus-nebula spectrum yields consistent stellar and nebular radial velocities, strongly supporting a physical association. Isochrone fitting to the color--magnitude diagram indicates a cluster age of $\sim90$\,Myr and a near-solar metallicity, implying a progenitor initial mass of $5.66^{+0.42}_{-0.37}\,M_{\odot}$. This value is among the highest empirical progenitor-mass constraints yet reported for any PN and approaches the lower boundary of the super-asymptotic giant branch (super-AGB) regime. We further find that the nebula is strongly nitrogen-enhanced, with an N/O ratio $\sim$7 times the solar value, broadly consistent with hot bottom burning in a relatively massive AGB progenitor. This system therefore provides a rare opportunity to test PN formation and nucleosynthesis at the high-mass end of the PN progenitor distribution.

\end{abstract}

\section{Introduction} \label{sec:intro}
Planetary nebulae (PNe) mark a short but important phase in the evolution of low- and intermediate-mass stars \citep[$\sim$0.8--8\,$M_\odot$;][]{Kwitter2022}. However, the progenitor properties of most PNe, especially their initial masses, remain difficult to determine directly. This hampers efforts to establish direct links between nebular properties and progenitor evolution, and hence to test how processes such as dredge-up, hot bottom burning (HBB), and mass loss shape the final outcomes of stellar evolution, as well as the broader astrophysical role of PNe in galactic ecosystems. The problem is particularly acute at the high-mass end of the PN progenitor distribution: although such systems are expected theoretically, observationally well-constrained systems remain extremely rare.

PNe associated with open clusters (OCs) are therefore particularly valuable, as the age and metallicity of the host cluster provide direct constraints on the properties of the progenitor star, especially its initial mass, which is generally difficult to determine for field PNe. The relatively young ages of OCs also make them especially useful for probing PNe descended from relatively massive progenitors. Unfortunately, such systems are exceedingly rare. In the Milky Way, numerous candidate PN--OC associations have been proposed over the past decades, and several have undergone repeated reassessment as improved photometric, kinematic, and astrometric constraints became available \citep[e.g.,][]{Majaess2007, Bonatto2008, MoniBidin2014, Frew2016, Gonzalez-Diaz2019}, leaving five comparatively convincing systems \citep{Parker2011,Fragkou2019a,Fragkou2019b,Fragkou2022a,Fragkou2022b,Werner2023,Fragkou2025,Bellini2025,Fragkou2026}. Notably, \citet{Fragkou2019b} reported an evolved PN associated with the $\sim$90\,Myr OC NGC\,6067, implying a progenitor mass of about $5.6\,M_{\odot}$ and thus representing one of the few well-constrained cases in the poorly sampled high-mass regime of PN progenitors, although the cluster membership of its central star still warrants further confirmation \citep{Bond2025}.

Outside the Milky Way, studies of PN--OC associations remain even more sparse. Early extragalactic work identified several candidate systems \citep{Larsen2006, Jacoby2013}, but in those cases either the emission-line sources could not be securely identified as bona fide PNe, or a chance superposition with the clusters could not be firmly ruled out. A major step forward came when \citet{Bond2015} identified a promising PN candidate in a relatively young OC in M31, which was subsequently established by \citet[hereafter D19]{Davis2019} as the first convincing extragalactic PN physically associated with a $290$\,Myr OC. More recently, \citet{Bond2025} reported another convincing PN--OC association in a $200$\,Myr cluster in the Large Magellanic Cloud (LMC).

Because secure PN--OC associations remain exceptionally rare, each new well-established system provides a valuable opportunity to link nebular properties to directly constrained progenitor characteristics. In this Letter, we report an extragalactic PN physically associated with a young OC in M31, whose age implies a progenitor initial mass of $\sim5.7\,M_{\odot}$, making it an unprecedented extragalactic anchor at the high-mass end of the PN progenitor distribution. Throughout this paper, we adopt a distance to M31 of $761\pm 11$\,kpc \citep{Li2021}, and a \citet{Cardelli1989} extinction law with $R_V=3.1$.

\section{Observations} \label{sec:observation}
\subsection{Photometric data}
We retrieve archival \textit{Hubble Space Telescope} (\textit{HST}) ACS/WFC images in F475W and F814W obtained as part of the Panchromatic Hubble Andromeda Treasury (PHAT) survey \citep{Dalcanton2012, Williams2014}, as well as additional ACS/WFC F625W and F658N images from program GO-14072 (PI: M. Boyer). The color-magnitude diagram (CMD) of the cluster region is constructed using the PHAT stellar photometric catalog \citep{Williams2023}, including only sources that passed all photometric quality cuts and had $\rm S/N>10$ in both F475W and F814W. Photometric uncertainties are estimated by interpolating the PHAT artificial star test results in magnitude and local stellar density. To minimize incompleteness effects, we further restrict the sample to stars at least 0.5 mag brighter than the local 50\% completeness limit \citep[Table~4 of][]{Williams2023}. To characterize the nebular [\ion{O}{3}] emission, we also retrieve narrowband [\ion{O}{3}] and $g$-band images from program 16BC25 (PI: M. Arnaboldi) through the CFHT MegaPrime archive\footnote{\url{https://www.cadc-ccda.hia-iha.nrc-cnrc.gc.ca/en/cfht/}}, and perform aperture photometry at the position of the source.

\subsection{Spectroscopy}
The source was observed with the Hectospec multifiber positioner and spectrograph on the 6.5\,m MMT telescope \citep{Fabricant2005} on 2011 September 26, as part of proposal 2011c-PA-11B-0083 (PI: P. Massey) targeting Wolf--Rayet stars in M31 \citep{Neugent2012, Neugent2014, Neugent2015}. The processed and calibrated spectrum is publicly available through the CFA Optical/Infrared Science Archive (OIRSA) Signature Programs\footnote{\url{https://oirsa.cfa.harvard.edu/signature_program/}}. It was obtained with the 270 gpm grating, providing wavelength coverage from 3650 to 9150\,\AA\ at a spectral resolution of $\sim5$\,\AA. The total exposure time was 3600\,s, consisting of three individual exposures of 1200\,s each. Although the MMT spectrum is flux calibrated, it is not on an absolute flux scale. We therefore rescale it by applying a constant multiplicative factor determined by matching the spectrum to the CFHT photometry. The transmission curves of corresponding filters are taken from the webpage of SVO filter profile services\footnote{\url{http://svo2.cab.inta-csic.es/theory/fps/}}. Although strong emission lines are detected in the spectrum, some caution is warranted because Hectospec is a fiber-fed instrument and the sky subtraction is therefore not based on a strictly local background measurement. For targets in the M31 disk, this raises the possibility that diffuse ionized gas from the galaxy disk may contaminate the observed spectrum. We discuss this potential effect further in Section~\ref{sec:association}.

\section{Establishing the OC--PN association} \label{sec:association}

Establishing a robust physical association between a Galactic PN and an OC typically requires consistency in projected position, distance, proper motion, reddening, and, most importantly, radial velocity. At the distance of M31, however, the available constraints are more limited: reliable proper-motion measurements are generally unavailable for individual sources, and whether the PN and the cluster are co-spatial along the line of sight cannot be determined directly. Consequently, the evidence for association relies primarily on their projected positional coincidence and radial-velocity consistency. We therefore focus on these two diagnostics below. 

\subsection{Evidence from Projected Position}
\begin{figure}
\centering 
\includegraphics[width=0.95\columnwidth]{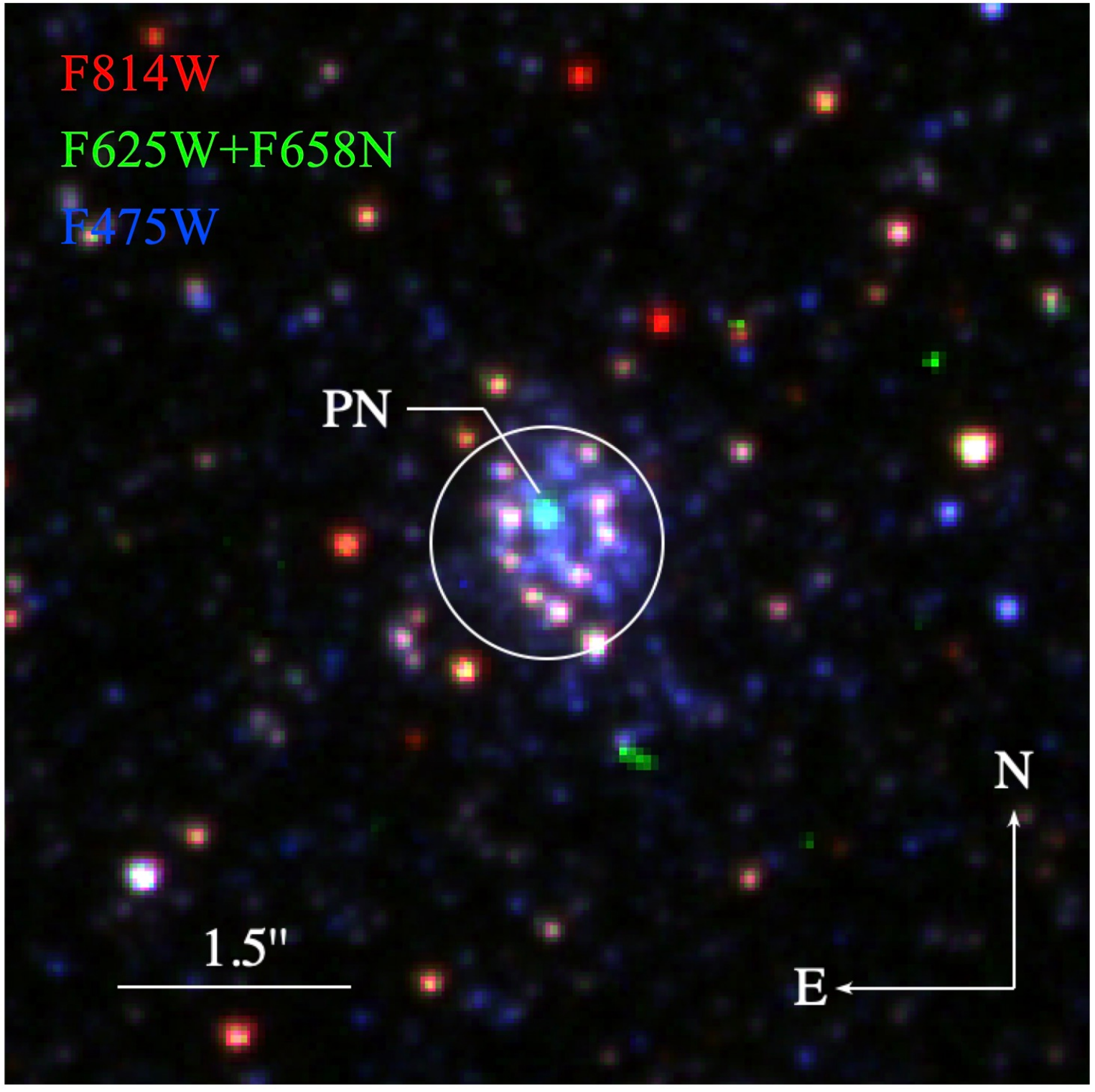}
\caption{
Color composite image of the young open cluster AP\,210, constructed from archival \textit{HST}/ACS observations in F475W (blue), F625W+F658N (green), and F814W (red). The PN appears as a bright blue-green point source near the cluster center because of its strong [\ion{O}{3}], H$\alpha$, and [\ion{N}{2}] emission. The white  circle indicates the $1\farcs5$ on-sky diameter of a single Hectospec fiber.
\label{fig:location}}
\end{figure}

The host cluster, centered at $\alpha=$\,00:45:39.17, $\delta=+$41:49:55.38 (J2000), lies on the northeastern segment of the 10\,kpc star-forming ring of M31. It was identified by the PHAT team as a young, relatively massive cluster based on high-resolution \textit{HST} data \citep{Johnson2015, Johnson2016}, with a derived age of $\log(\mathrm{Age/yr})=8.0$ and a mass of $\log(M/M_{\odot})=4.15$. Following their naming convention, we refer to it throughout this paper as AP\,210. Figure~\ref{fig:location} shows a color composite of the cluster constructed from the F475W, F625W, F814W, and F658N images. The cluster appears compact and slightly elongated, with a ring-like distribution of relatively bright stars around its outskirts. Although the PN candidate identified by \citet{Merrett2006} from [\ion{O}{3}] $\lambda$5007 emission, designated M\,279 in their catalog, is not obvious in the PHAT broadband images alone, the inclusion of the F658N image, which covers H$\alpha$ and [\ion{N}{2}] emission, reveals a bright point source near the cluster center. The source lies well within the effective radius of $0\farcs53$ measured for AP\,210 by \citet{Johnson2015}, providing strong evidence for a close projected positional association. The presence of a compact emission source in the F658N image, together with the clean local background and the lack of spatially extended emission features, suggests that the emission lines detected in our spectrum arise from this point source rather than from diffuse ionized gas in the M31 disk.

This close positional agreement favors a physical association, but it cannot by itself exclude a chance superposition. D19 estimated a probability of only $\sim$1\% for such an alignment to occur at random. We revisit this question using a more complete sample of young clusters and PNe within the PHAT footprint, where both populations are comparatively well characterized. In this region, 957 young clusters with ages of 50\,Myr--1\,Gyr are cataloged by \citet{Johnson2016}, together with 700 likely PNe identified by \citet{Bhattacharya2019}. As a spatial prior for the PN population, we adopt the smoothed density distribution of RGB stars selected from the PHAT catalog following \citet{Williams2023}, using the RGB stars as a proxy for the underlying stellar density traced by the parent population of PNe. We then perform Monte Carlo (MC) simulations in which each realization contains $N_{\rm PN}=700$ synthetic PNe, and we count the number of overlaps with clusters, where an overlap is defined as a mock PN falling within one effective radius of a cluster center. Cluster effective radii are taken from \citet{Johnson2015}, and the PNe are assumed to be point sources. The resulting empirical probability of finding at least one chance superposition within the PHAT footprint is 6.64\%, increasing to 9.12\% when the additional 292 possible PNe from \citet{Bhattacharya2019} are also included.

Although this probability is modest, it is not negligible. Our simulation also ignores possible age-dependent differences in the spatial distribution of PNe. PNe descended from relatively massive progenitors may more closely trace the young stellar component and spiral structure of M31, making them more likely to coincide in projection with young clusters. However, such high-mass PNe are intrinsically rare and may also remain visible for relatively short times, so the net effect on the chance-superposition probability is not straightforward to predict. Positional agreement alone is therefore insufficient to exclude a chance superposition, and radial-velocity consistency provides a critical additional test.

\begin{figure*}[t]
\centering 
\includegraphics[width=2.0\columnwidth]{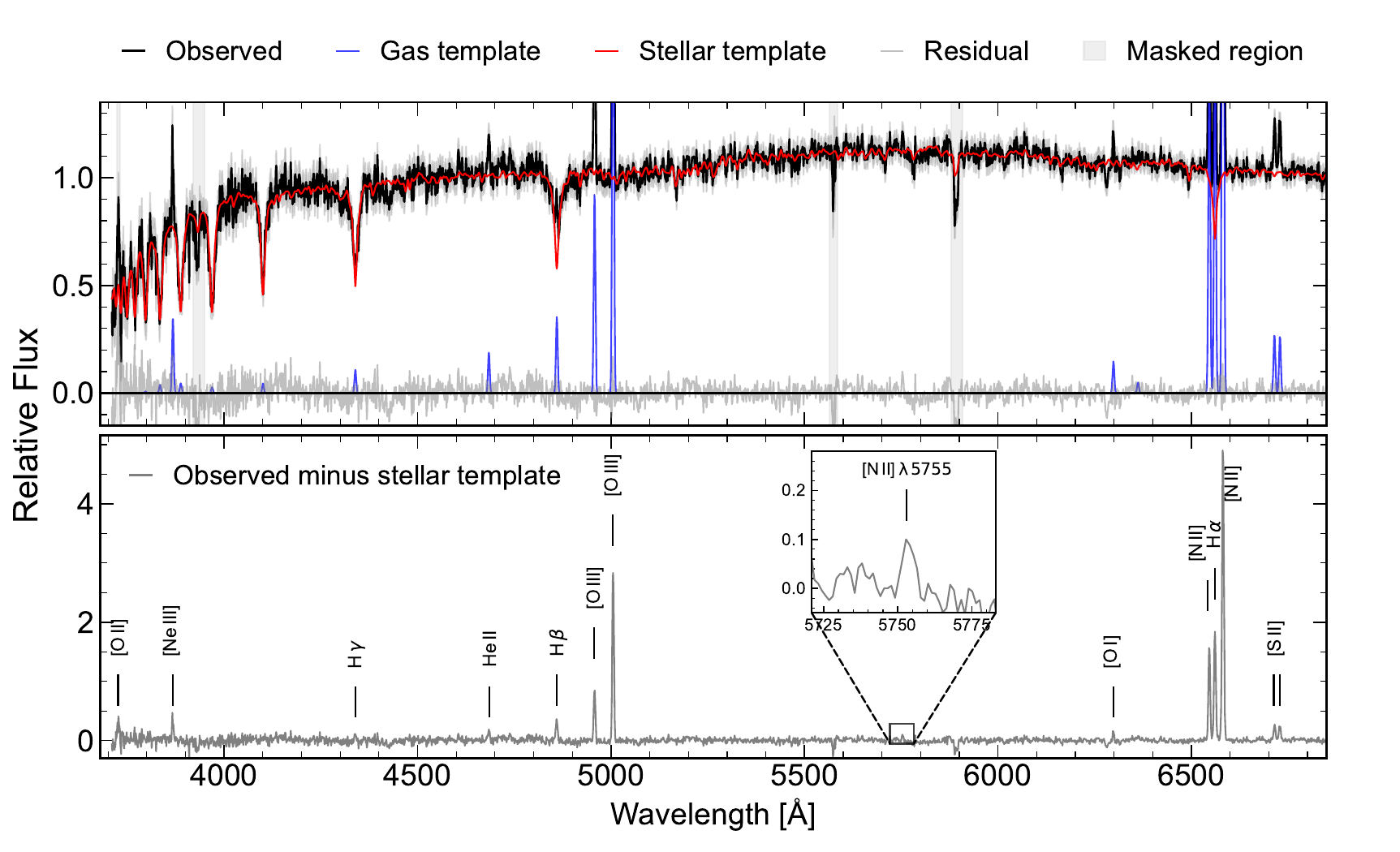}
\caption{
Top panel: observed spectrum (black), best-fit stellar spectrum (red), and best-fit nebular-emission model (blue) from the pPXF fit. The fit residuals are shown in gray, and masked regions are indicated by gray shading. Bottom panel: residual nebular spectrum obtained by subtracting the best-fit stellar spectrum from the observed spectrum. Key nebular emission lines used in the subsequent analysis are labeled.
\label{fig:spec}}
\end{figure*}

\subsection{Evidence from Radial Velocity}\label{subsec:rv}
Radial velocity provides a decisive test of the OC--PN association because OCs typically have very small internal velocity dispersions \citep[$\sim0.5$--$3\,\kms$; e.g.,][]{Mathieu2000, Geller2010}. For our source, however, this test is observationally  challenging because the PN is located very close to the cluster center. Consequently, the observed spectrum inevitably contains a blend of PN and cluster light rather than two cleanly separable components (see Figure~\ref{fig:location}). We therefore use the penalized pixel-fitting code \citep[pPXF;][]{Cappellari2004, Cappellari2017, Cappellari2023}, which performs full spectrum fitting to model the stellar continuum and nebular emission lines simultaneously as distinct kinematic components. Similar approaches have been successfully applied to separate PN emission from the galaxy background continuum \citep[e.g.,][]{Roth2021, Soemitro2025}. 

We apply pPXF to the MMT spectrum over the wavelength range 3710--8000\,\AA\ to avoid possible second-order contamination at longer wavelengths, masking regions affected by interstellar absorption (e.g., the \ion{Na}{1}\,D lines). We adopt the E-MILES stellar population synthesis models of \citet{Vazdekis2016} and select the nine templates with ages of 0.0794, 0.1000, and 0.1259\,Gyr and metallicities of [M/H] = $-0.40$, 0.00, and 0.22, guided by our CMD-fitting results (Section~\ref{subsec:progenitor}). For the nebular emission, we assign two separate kinematic components: one for the high-excitation lines ([\ion{Ne}{3}] $\lambda$3869, \ion{He}{2} $\lambda$4686, and [\ion{O}{3}] $\lambda\lambda$4959,5007), and the other for the Balmer lines and low-excitation forbidden lines (e.g., [\ion{N}{2}] and [\ion{S}{2}]). During the fit, we include a 12th-order Legendre multiplicative polynomial to adjust the continuum shape of the model spectrum, accounting for template mismatch, flux-calibration uncertainties, and reddening-induced continuum variations. No additive polynomial is used, since it can modify absorption line strengths and bias the derived stellar population parameters. We estimate the uncertainties of the derived parameters using a wild bootstrap analysis \citep{Wu1986, Davidson2008}. Specifically, we resample the residuals from the initial fit with random Rademacher weights and refit the spectrum 1000 times.

The fitting results are shown in Figure~\ref{fig:spec}. We measure radial velocities of $-87.6 \pm 5.6\,\kms$ for the cluster component, $-92.0 \pm 2.4\,\kms$ for the high-excitation emission-line component, and $-86.5 \pm 0.9\,\kms$ for the Balmer and low-excitation emission-line component. A modest velocity offset between the two emission-line components is not unexpected, as lines of different excitation may arise in different ionization structures within the nebula and thus trace different gas kinematics. Within the uncertainties, both emission-line components are fully consistent with the cluster velocity, providing strong evidence for a physical association between the OC and PN. For comparison, the surrounding PN population in the \citet{Merrett2006} catalog has a mean radial velocity of $\sim-110\,\kms$ and a velocity dispersion of $\sim60\,\kms$. The observed PN--cluster velocity offsets of only a few $\kms$ are therefore small compared to the characteristic velocity scale of the local PN population. Combined with the close projected positional coincidence, these results lead us to conclude that the PN is physically associated with AP\,210.

The pPXF fit also provides luminosity-weighted estimates of the cluster age and metallicity, yielding \(99.8^{+3.3}_{-6.0}\,\)Myr and [M/H]\(=0.00^{+0.05}_{-0.04}\), respectively. These values are consistent with our CMD-based results within the uncertainties (see Section~\ref{subsec:progenitor}). We do not use the pPXF-based estimates in the subsequent analysis, however, because suitable templates for such a young cluster are sparse, making the fit more vulnerable to parameter degeneracies \citep[e.g.,][]{Chen2016}. 

\section{A PN from a High-Mass Progenitor} \label{sec:discussion}
\subsection{Progenitor-mass Constraint from the Host Cluster}\label{subsec:progenitor}
The age and metallicity of the host cluster provide direct constraints on the properties of the PN progenitor, especially its initial mass. We therefore fit the CMD of AP\,210 with theoretical isochrones. The CMD is constructed from the resolved stellar photometry catalog of \citet{Williams2023}. Stars within the cluster photometric aperture \citep[$R_{\rm ap}=1\farcs77$;][]{Johnson2015} are treated as candidate members, while the local background population is characterized using an annulus with inner and outer radii of $1.2R_{\rm ap}$ and $3.2R_{\rm ap}$, respectively. We adopt a field-decontamination method similar to that of D19. In each realization, control-field stars are randomly sampled according to the area ratio between the cluster and control regions. For each sampled field star, we identify its nearest neighbor among stars in the cluster aperture using a CMD-space distance metric $s$ (see D19 for details). We then remove the matched cluster star if $s \le s_{\rm lim}$, where $s_{\rm lim}$ is the adopted threshold for accepting the match as a plausible field-star counterpart. Repeating this procedure over 500 realizations, we define the empirical retention fraction \(P_{\rm mem}\) of each star as the fraction of realizations in which it is retained, and keep only stars with \(P_{\rm mem}\ge P_{\rm lim}\). 

We then fit the decontaminated CMD with theoretical isochrones from the MESA Isochrones and Stellar Tracks \citep[MIST;][]{Dotter2016, Choi2016}, adopting a stellar rotation of \(v/v_{\rm crit}=0.4\). Reddening is applied through the MIST bolometric-correction tables. The fit is performed using the \texttt{emcee} implementation of the Markov Chain Monte Carlo \citep[MCMC;][]{Foreman-Mackey2013}, with age, metallicity ([Fe/H]), reddening, and distance modulus as free parameters. We adopt a Gaussian prior on the distance modulus of M31 based on \citet{Li2021}, and uniform priors on all other parameters. Because bright evolved stars can dominate the likelihood owing to their very small photometric uncertainties, we apply a down-weighting factor \(f_{\rm down}\) to these stars.

\begin{figure}
\centering 
\includegraphics[width=1.0\columnwidth]{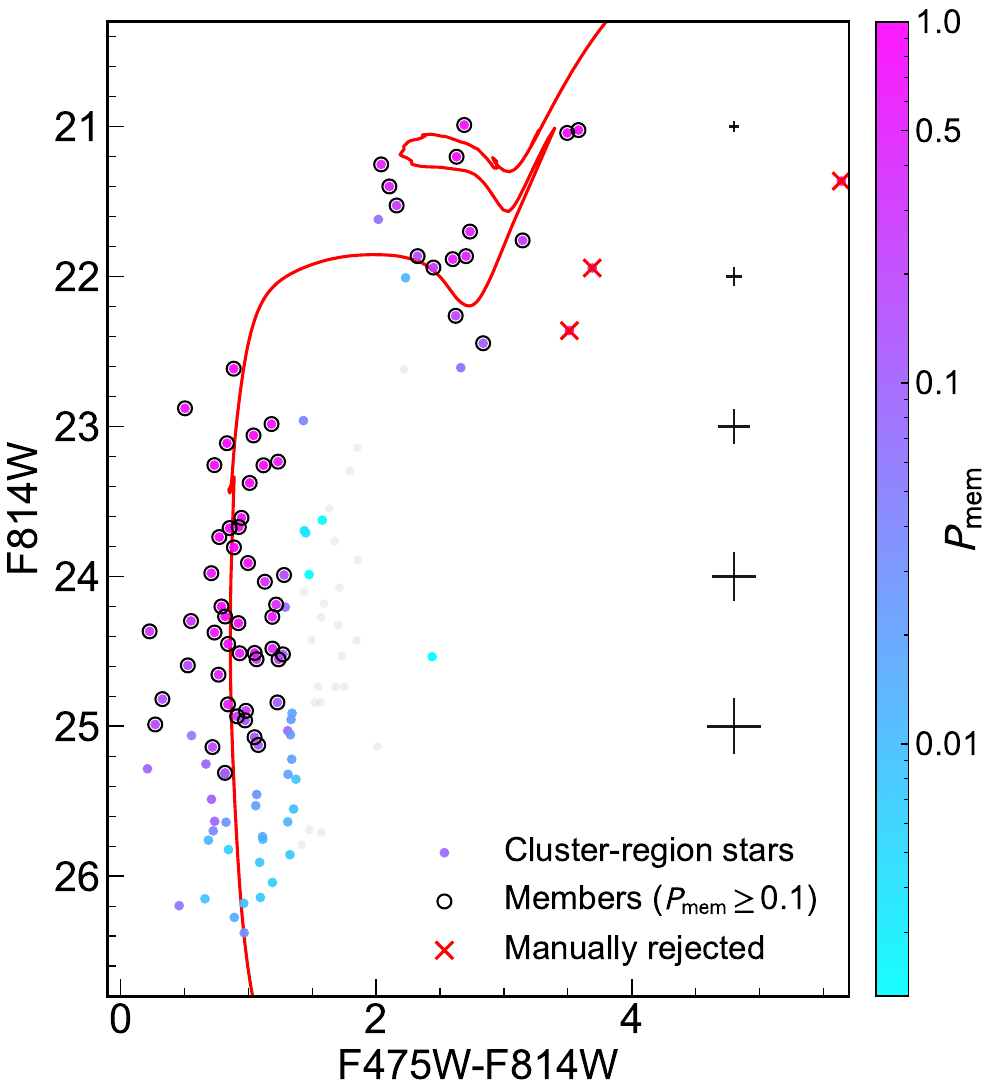}
\caption{
Color--magnitude diagram of AP\,210. Stars within the cluster aperture are color-coded by their empirical retention fraction (\(P_{\rm mem}\)), with stars of \(P_{\rm mem}=0\) shown in light gray. Black open circles mark stars with \(P_{\rm mem}>0.1\) retained for the isochrone fitting, while red crosses indicate outliers excluded by visual inspection. The red curve shows the best-fitting MIST isochrone, with \(\log(\mathrm{Age/yr})=7.95\), \(E(B-V)=0.57\)\,mag, and \([\mathrm{Fe/H}]=0.06\). Representative photometric uncertainties as a function of magnitude are shown on the right.
\label{fig:cmd}}
\end{figure}

For our fiducial fit, we adopt \(s_{\rm lim}=0.55\), \(P_{\rm lim}=0.1\), and \(f_{\rm down}=0.1\), and exclude three obvious outliers by visual inspection. The MCMC fit yields cluster parameters of \(\log(\mathrm{Age/yr})=7.95\pm 0.05\), \([\mathrm{Fe/H}]=0.06\pm0.08\), and \(E(B-V)=0.57\pm0.02\)\,mag. Figure~\ref{fig:cmd} shows the CMD of AP\,210 together with the best-fit isochrone. Although the main sequence of AP210 is well detected and provides strong constraints on age and reddening, a significant age--metallicity degeneracy remains. Moreover, because of the  large reddening, giant-branch members are difficult to distinguish reliably from field stars in the CMD, even though this region would otherwise provide important leverage on age and metallicity. We therefore repeat the analysis over a broad range of choices for \(s_{\rm lim}\), \(P_{\rm lim}\), and \(f_{\rm down}\) in order to quantify the systematic uncertainties introduced by member selection and the weighting assigned to bright evolved stars. As a result, we adopt final cluster parameters for AP210 of an age of \(90\pm15\)\,Myr, a metallicity of \(\rm [Fe/H]=0.06\pm0.10\), and a reddening of \(E(B-V)=0.57\pm0.02\)\,mag. These values are consistent with the CMD-based results of \(\log(\mathrm{Age/yr})=8.0\) and \(A_V=1.8\)\,mag reported by the PHAT team \citep{Johnson2016, Johnson2022}, although those authors used the MATCH software \citep{Dolphin2002} and PARSEC-COLIBRI isochrones \citep{Bressan2012, Marigo2017}.

Based on the derived physical parameters of AP\,210 and the MIST models, we infer an initial mass of \(5.66^{+0.42}_{-0.37}\,M_{\odot}\) for the PN progenitor and a final mass of \(0.94^{+0.04}_{-0.03}\,M_{\odot}\) for the post-asymptotic giant branch (post-AGB) remnant. This places it among the most massive PN progenitors yet constrained with comparable precision and provides a rare extragalactic observational anchor at the high-mass end of the PN progenitor distribution, analogous to the cluster PN studied by \citet{Fragkou2019b}. Both systems lie close to the lower-mass boundary of the super-AGB regime \citep[$\sim$6--12\,$M_{\odot}$][]{Doherty2017}. Note that the mass inference assumes standard single-star evolution.  Although binary interaction could in principle modify the mapping between cluster age and progenitor mass, there is no direct evidence that such effects are required in this system.

\subsection{Nebular Properties and Nitrogen Enhancement}

\begin{deluxetable}{lccc}
\tablecaption{Measured ($F$) and dereddened ($I$) line fluxes.
\label{tab:linefluxes}}
\tablewidth{0pt}
\tablehead{
\colhead{Ion} &
\colhead{Wavelength (\AA)} &
\colhead{$F\pm \Delta F$} &
\colhead{$I\pm \Delta I$}
}
\startdata
{[\ion{O}{2}]}  & 3727+3729 & $136 \pm 23$   & $251 \pm 60$ \\
{[\ion{Ne}{3}]} & 3869      & $81.1 \pm 8.5$ & $140 \pm 25$ \\
\ion{H}{1} & 4340 & $40.2 \pm 8.7$ & $54.1 \pm 12.6$ \\
\ion{He}{2}     & 4686      & $48.8 \pm 9.0$ & $53.8 \pm 10.2$ \\
\ion{H}{1}      & 4861      & 100            & 100 \\
{[\ion{O}{3}]}  & 4959      & $223 \pm 21$   & $212 \pm 18$ \\
{[\ion{O}{3}]}  & 5007      & $723 \pm 60$   & $673 \pm 48$ \\
{[\ion{N}{2}]}  & 5755      & $23.3 \pm 7.8$ & $16.5 \pm 5.2$ \\
{[\ion{N}{2}]}  & 6548      & $428 \pm 38$   & $243.8 \pm 8.6$ \\
\ion{H}{1}      & 6563      & $498 \pm 44$   & $282.8 \pm 2.9$ \\
{[\ion{N}{2}]}  & 6583      & $1327 \pm 112$ & $750 \pm 23$ \\
{[\ion{S}{2}]}  & 6716      & $75.6 \pm 12.2$ & $41.5 \pm 5.8$ \\
{[\ion{S}{2}]}  & 6731      & $75.4 \pm 12.8$ & $40.8 \pm 6.1$ \\
$c(\mathrm{H}\beta)$ & \multicolumn{3}{c}{$0.83^{+0.14}_{-0.12}$} \\
\enddata
\tablecomments{Fluxes are normalized to H$\beta=100$.}
\end{deluxetable}

We derive the nebular spectrum by subtracting the best-fit stellar spectrum from the observed spectrum. This subtraction removes the stellar continuum but does not deredden the nebular emission. The resulting spectrum therefore represents the observed, attenuated PN spectrum. We compute the corresponding flux uncertainties by adding in quadrature the per-pixel observational errors and the additional uncertainty estimated from the wild bootstrap procedure. As shown in the bottom panel of Figure~\ref{fig:spec}, a striking feature of the nebular spectrum is the exceptional strength of the [\ion{N}{2}] lines, with [\ion{N}{2}] \(\lambda 6583\) nearly three times stronger than H\(\alpha\). This strongly suggests that the PN is nitrogen enhanced and likely of Type~I chemistry \citep{Peimbert1978, Peimbert1983, Torres-Peimbert1997}. In addition, the detection of \ion{He}{2} \(\lambda4686\) indicates that this is a high-excitation PN with a central star effective temperature exceeding 50\,kK. 

We measure line fluxes by fitting Gaussian profiles together with a linear function for the local continuum. In regions where lines are blended, multiple Gaussian components are fitted simultaneously. We then perform plasma diagnostics with the \texttt{PyNeb} package \citep{Luridiana2015}, version 1.1.19. The electron temperature ($T_e$) and density ($n_e$) are derived from the [\ion{N}{2}] $(\lambda 6548+\lambda 6583)/\lambda 5755$ ratio and the [\ion{S}{2}] $\lambda 6716/\lambda 6731$ ratio, respectively. The extinction coefficient, $c(\mathrm{H}\beta)$, is determined by comparing the observed and theoretical H$\alpha$/H$\beta$ ratios. In practice, we solve for $c(\mathrm{H}\beta)$, $T_e$, and $n_e$ iteratively, starting from a standard Case B assumption of $T_e=10{,}000$\,K and $n_e=1000$\,cm$^{-3}$, until the solution converges. Uncertainties are estimated from 1000 MC realizations by propagating the line-flux errors. We obtain $T_e=11,814^{+1792}_{-1900}$\,K, $n_e=711^{+744}_{-451}\,\mathrm{cm^{-3}}$, and $c(\mathrm{H}\beta)=0.83^{+0.14}_{-0.12}$. The measured and dereddened line fluxes are listed in Table~\ref{tab:linefluxes}. We defer further discussion of the extinction value to Section~\ref{subsec:extinction}.

Accurate elemental abundances are difficult to derive from the available data. In particular, no [\ion{O}{3}]-based estimate of \(T_e\) is available, even though it is especially important for determining the ionic abundances of high-ionization ions. We therefore compute ionic abundances of O$^+$/H$^+$ and N$^+$/H$^+$ using the [\ion{N}{2}]-based $T_e$, with the caveat that  it may be biased by high-density clumps within the PN that can suppress the [\ion{N}{2}] $\lambda\lambda 6548, 6583$ nebular lines via collisional de-excitation \citep{Morisset2017}, and by recombination excitation contributing to the [\ion{N}{2}] $\lambda 5755$ auroral line \citep{Liu2000}. Because no \ion{He}{1} line is detected, we cannot constrain the ionization correction factor (ICF) required to derive O/H. This also prevents a reliable estimate of the ICF needed for N/H, which depends on O/H. Nevertheless, the N/O ratio, which encodes crucial information about the PN progenitor, can still be estimated under the classical assumption N/O=N$^{+}$/O${^+}$ \citep{Peimbert1969, Kingsburgh1994}. This approximation is known to break down in highly ionized PNe, i.e., when $\omega={\rm O}^{++}/({\rm O}^+ +{\rm O}^{++})>0.9$ \citep{Delgado-Inglada2014, Morisset2023}. Since our object does not fall in this extreme high-\(\omega\) regime, we adopt it here as a reasonable approximation.

We measure an N/O ratio of 0.98$^{+0.31}_{-0.27}$ for the PN, corresponding to $\log$(N/O)$=-0.01^{+0.12}_{-0.14}$. This confirms that the nebula is strongly nitrogen enriched and is consistent with a relatively massive progenitor that underwent HBB during the AGB phase, when CNO-cycle products are efficiently mixed to the surface and later expelled through mass loss \citep[e.g.,][]{Karakas2014, Karakas2016, Ventura2017, Henry2018}. However, the measured N/O is higher than predicted by the solar-metallicity models of \citet{Karakas2016} for a progenitor of \(\sim5.7\,M_{\odot}\), and is instead more consistent with their super-solar metallicity model (\(Z=0.03\)). Finally, although the high N/O places the PN in the Type~I regime \citep[$\log(\mathrm{N/O})\geq-0.3$;][]{Peimbert1983}, a secure classification still requires a reliable measurement of the helium abundance.

\subsection{Extinction and [\ion{O}{3}] Luminosity}\label{subsec:extinction}

We measure an [\ion{O}{3}] magnitude of \(m_{5007}=23.67\)\,mag for the PN, where \(m_{5007}=-2.5\log F_{5007}-13.74\) and \(F_{5007}\) is the flux of the [\ion{O}{3}] \(\lambda5007\) line in erg\,cm\(^{-2}\)\,s\(^{-1}\) \citep{Jacoby1989}. This value is slightly fainter than the \(m_{5007}=23.39\)\,mag reported by \citet{Merrett2006}, likely because the earlier measurement included contamination from the superposed cluster light, as noted by D19. Based on the Balmer decrement, we measure an extinction coefficient of $c(\mathrm{H}\beta)=0.83^{+0.14}_{-0.12}$. Correcting for this extinction yields \(m_{5007,\mathrm{dered}}=21.69\)\,mag, corresponding to \(M_{5007}=-2.72\)\,mag at the distance of M31. The PN therefore lies roughly 2\,mag below the bright-end cutoff of the planetary nebula luminosity function (PNLF), for which \(M^*\approx-4.5\)\,mag \citep{Jacoby1989}, implying that it is not currently at the evolutionary stage of maximum [\ion{O}{3}] output, when its intrinsic \(M_{5007}\) could significantly exceed \(M^*\). Assuming a maximum [\ion{O}{3}] $\lambda5007$ conversion efficiency of 12\% \citep[e.g.,][]{Dopita1992, Schonberner2010}, the observed [\ion{O}{3}] luminosity implies a lower limit of \(\sim10^3\,L_{\odot}\) for the luminosity of the central star.

Intriguingly, the extinction coefficient measured from the Balmer decrement translates into $E(B-V)=0.57^{+0.10}_{-0.09}$\,mag. This value is consistent with the extinction of $E(B-V)=0.57^{+0.02}_{-0.02}$\,mag measured for the host cluster from CMD fitting, albeit with relatively large uncertainty. If we assume that the PN experiences the same line-of-sight reddening as the cluster stars, this agreement suggests that no substantial additional circumnebular reddening is detected toward the PN. This is somewhat unexpected for a PN that is still powered by a central star with a luminosity of at least $\sim10^3\,L_{\odot}$, because for a $5.7\,M_{\odot}$ progenitor the post-AGB evolution is extremely rapid \citep[e.g.,][]{Miller2016, Gesicki2018}, and the gas and dust ejected during the AGB phase are not expected to have fully dispersed. For further discussion of the relationship between central-star core mass and circumnebular extinction, and of their influence on the observed PNLF, see \citet{Davis2018} and \citet{Jacoby2025}.

\begin{figure}
\centering 
\includegraphics[width=1.\columnwidth]{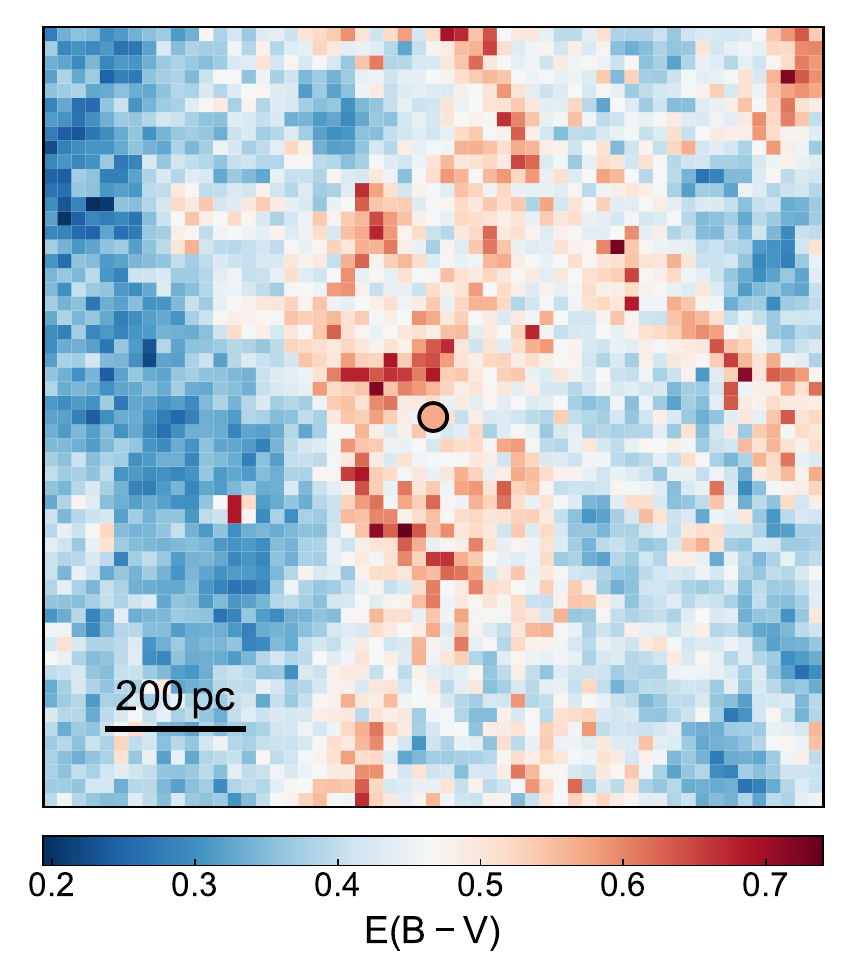}
\caption{
Location of AP\,210 overlaid on a high-resolution extinction map of M31 (M.X. Sun et al. 2026, in preparation). North is up and east is to the left. The cluster position is marked by a circle at the center of the panel, with its color corresponding to the measured extinction of $E(B-V)=0.57$\,mag.\label{fig:env}}
\end{figure}

We further compare this extinction value with that of its surrounding environment. As shown in Figure~\ref{fig:env}, AP\,210 lies on the $\sim10$\,kpc star-forming ring of M31, where the local extinction is relatively high but also shows noticeable spatial variation. The extinction value of $E(B-V)=0.57$\,mag is close to the characteristic extinction of its immediate surroundings, consistent with the system being embedded in a dusty spiral-arm environment. One plausible explanation for the lack of strong additional circumnebular reddening is that the PN has a bipolar or otherwise axisymmetric morphology, as is often seen in Galactic Type~I PNe. If the dust is concentrated in an equatorial waist or torus, the inferred reddening can depend strongly on viewing angle, and a more pole-on geometry would yield substantially smaller reddening than an edge-on view \citep[e.g.,][]{MoragaBaez2023, Pignata2024, Smith-Perez2026}. The complex dust structure of the spiral arm, together with potentially patchy extinction within the young massive cluster, may further blur any difference between the extinction toward the PN and that affecting the cluster stars. Thus, while the current extinction measurement does not reveal a strong additional circumnebular reddening component, it does not by itself exclude the presence of circumnebular dust.

\section{Summary} \label{sec:summary}

In this Letter, we report a PN physically associated with a young OC in M31, adding a third compelling case to the very small sample of known extragalactic PN--OC systems. High-resolution \textit{HST} broadband imaging, together with the narrowband  image that reveals the nebular emission, shows that the PN lies nearly at the cluster center. By decomposing the observed spectrum into stellar and nebular components, we simultaneously measure the radial velocities of the cluster and the PN, and find them to be consistent within the uncertainties. Together, these positional and kinematic lines of evidence strongly support a physical association between the PN and the OC. Isochrone fitting to the cluster CMD yields an age of $90\pm15$\,Myr and a metallicity of $\rm [Fe/H]=0.06\pm0.10$. This implies a progenitor initial mass of $5.66^{+0.42}_{-0.37}\,M_{\odot}$, placing it among the PNe with the highest progenitor masses constrained with reasonable precision. The nebular spectrum shows strong nitrogen enhancement, with N/O$=0.98^{+0.31}_{-0.27}$, broadly consistent with HBB in a relatively massive AGB progenitor, although not fully reproduced by current theoretical models. The extinction inferred from the Balmer decrement is consistent with  the cluster reddening, suggesting no clear evidence for strong additional circumnebular reddening, although circumnebular dust cannot be excluded. This system provides an unprecedented extragalactic observational anchor near the theoretically expected upper end of the PN progenitor distribution, showing that stars with initial masses of nearly $5.7\,M_{\odot}$ can produce a visible PN while also offering a rare opportunity to test nucleosynthesis in a relatively massive AGB progenitor.

\section*{Acknowledgments}
We are grateful to the anonymous referee, whose excellent comments and suggestions significantly improved this article.
P.C. thanks Mingxu Sun for sharing the latest extinction map of M31. P.C. also thanks Yuxin Ke and Jiyu Wang for helpful discussions. This work is partially supported by the National Natural Science Foundation of China 12173034, 12322304, 12322306, 12173047, and 12373028, the National Natural Science Foundation of Yunnan Province 202301AV070002 and the Xingdian talent support program of Yunnan Province. We acknowledge the science research grants from the China Manned Space Project with no. CMS-CSST-2025-A11. X.C. and S.W. acknowledge support from the Youth Innovation Promotion Association of the Chinese Academy of Sciences (Nos. 2022055 and 2023065). We gratefully acknowledge the use of publicly available data obtained with the 6.5\,m MMT through the CfA Optical/Infrared Science Archive (OIRSA) Signature Programs.
Based on observations obtained with MegaPrime/ MegaCam, a joint project of CFHT and CEA/DAPNIA, at the Canada-France-Hawaii Telescope (CFHT) which is operated by the National Research Council (NRC) of Canada, the Institut National des Sciences de l'Univers of the Centre National de la Recherche Scientifique of France, and the University of Hawaii. This research used the facilities of the Canadian Astronomy Data Centre operated by the National Research Council of Canada with the support of the Canadian Space Agency. Some of the data presented in this paper were obtained from the Mikulski Archive for Space Telescopes (MAST) at the Space Telescope Science Institute. The specific observations analyzed can be accessed via \dataset[https://doi.org/10.17909/ergb-jg63]{https://doi.org/10.17909/ergb-jg63}. STScI is operated by the Association of Universities for Research in Astronomy, Inc., under NASA contract NAS5-26555. Support to MAST for these data is provided by the NASA Office of Space Science via grant NAG5-7584 and by other grants and contracts.

\bibliography{sample701.bib}{}
\bibliographystyle{aasjournalv7}

\end{document}